\documentclass{iauc}
\usepackage{graphicx}

\title[The progenitor of Canis Major dwarf] 
{The Canis Major dwarf galaxy as the progenitor of the Monoceros tidal stream.}

\author[Mart\'\i nez-Delgado et al.]   
{D. Mart\'\i nez-Delgado$^1$%
  \thanks{email: ddelgado@iaa.es},
J. Pe\~narrubia$^2$, D. I. Dinescu$^3$, D.J. Butler$^2$ \break and H. W. Rix$^2$}

\affiliation{$^1$Instituto de Astrof\'\i sica de Andaluc\'\i a (CSIC), Spain.
$^2$ Max-Planck-Institut-fur Astronomie, Germany. $^3$ Astronomy Department, Yale University, USA.}

\pubyear{2005}
\volume{198} 
\pagerange{119--126}
\date{first draft}
\jname{Near Field Cosmology with Dwarf Elliptical Galaxies }
\editors{Helmut Jerjen \& Bruno Binggeli, eds.}
\begin{document}

\maketitle

\begin{abstract}

The Sloan Digital Sky Survey has recently discovered a
coherent ring of  stars at low galactic latitude that is believed to be
the tidal stream of  a merging dwarf galaxy in the Galactic plane
 (named the Monoceros tidal stream). The existence and location of the core of its progenitor galaxy is still controversial. The best candidate is the 
Canis Major dwarf galaxy, a distinct overdensity of red stars
 discovered in the 2MASS survey, but also interpreted as the signature of
 the Galactic warp  viewed in projection. In this paper, we report 
a variety of new observational evidence that supports the notion that
  CMa is the remnant of a partially disrupted core of a dwarf satellite. The comparison of the orbit 
derived from our theoretical model for the parent galaxy of this ring-like structure with an accurate determination of CMa 
orbit leads to the conclusion that this 
satellite  is the best candidate for the progenitor of the Monoceros tidal stream


\end{abstract}

\firstsection 
\section{Introduction}

The Sloan Digital Sky Survey has recently discovered a
coherent ring-like structure at low galactic latitude spanning about 100
degree in the sky and surrounding the Galactic disk at Galactocentric
distances from $\sim$ 15 kpc to $\sim$ 20 kpc (Newberg et al. 2002; Yanny
et al. 2003). The tremendous observational and theoretical effort to understand its origin have
proven that the structural characteristic and kinematics of this stellar ring are consistent with the 
properties expected for the tidal stream of a merging
 dwarf galaxy in the Galactic plane (named the Monoceros tidal stream). If this
ring is a tidal tail feature, it must have had a parent galaxy, which may or
may not be completely disrupted by now.

Unlike the Sagittarius tidal stream, the Monoceros stream has been detected
prior to locating the main body of its progenitor galaxy. The best available candidate is the Canis Major (CMa) dwarf galaxy, a strong 
elliptically shaped stellar overdensity of red giant stars
discovered by Martin et al. (2004) from an analysis of the 2MASS survey.
As an alternative interpretation, Momany et al. (2004) suggested that this
overdensity is only the signature of the Galactic warp. In this paper, we address the controversy on the origin of this stellar system and its relation with
the Monoceros stream by answering the following questions: Can we constrain the position of the progenitor galaxy with N-body simulations from the distribution and kinematics of Monoceros tidal debris? Is the CMa over-density the remmnat of the core of a dwarf galaxy? And if it is, is CMa the progenitor of the Monoceros tidal stream?

\begin{figure}
\centering
 \includegraphics{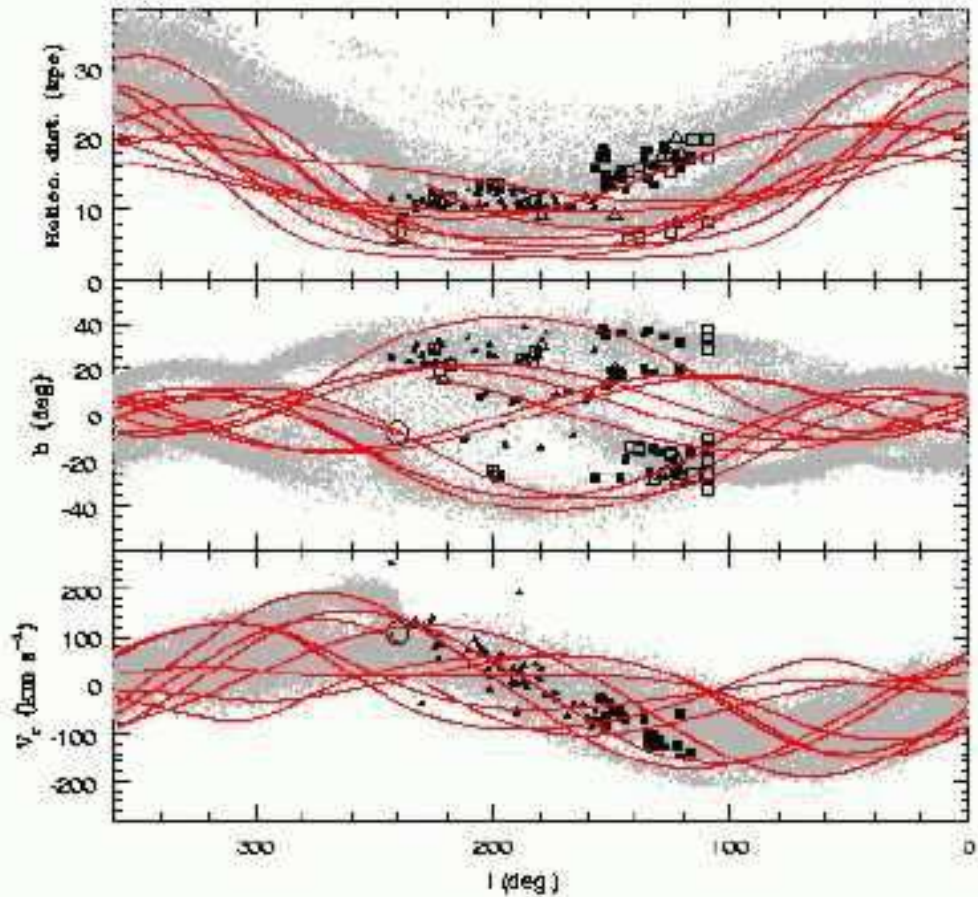}
  \caption{Comparison of the best fitting N-body simulation of the Monoceros
tidal stream (small dots;see Pe\~narrubia et al. 2005 and references therein
for a complete description of the Monoceros stream detections) with the
orbit of the CMa dwarf derived from its absolute proper motion (solid line) by
 Dinescu et al. (2005)}\label{fig:contour}
\end{figure}

\section{Constraining the progenitor position with theoretical model of the Monoceros tidal stream}

 Pe\~narrubia et al. (2005) have attempted to determine the present position 
of the Monoceros stream progenitor through N-body simulations. These authors 
constrain the motion and mass of a possible progenitor by fitting the 
available Monoceros stream detections to thousands of orbits of dwarf galaxies 
with different masses. This method has been proven to provide powerful 
constrains on the eccentricity and orbital inclination of the possible 
progenitor, finding that the best solutions are for a satellite galaxy 
moving on a low eccentric (e=0.10 $\pm$ 0.05), low orbital inclination 
(i=25 $\pm$ 5 deg) prograde orbit (Figure 1).
Owing to the small area of the sky where the Monoceros stream has been 
detected, the solutions are degenerated for some free parameters of the model: 
Namely, the axis-ratio of the Milky Way halo, the mass and the present 
location of the stream progenitor cannot be sufficiently constrained.   
Focusing on the last point, the model reproduces the geometrical and 
kinematical distribution of debris if the progenitor's remnants are 
located within $260>l>120$ at heliocentric distance of $\sim$14 kpc.
 That solution can be futher constrained with future detections in a 
wider range of longitudinal directions.
 The CMa system  is located at l=240, which enters in the range obtained from 
N-body simulations, although at a closer distance ($d_{helioc}\sim$8 kpc).
The close distance of CMa appears to indicate that, if this system is the
 progenitor of the Monoceros stream,  there are parts of the tidal stream at closer heliocentric distances that 
 have not yet been detected.

\begin{figure}
\centering
 \includegraphics{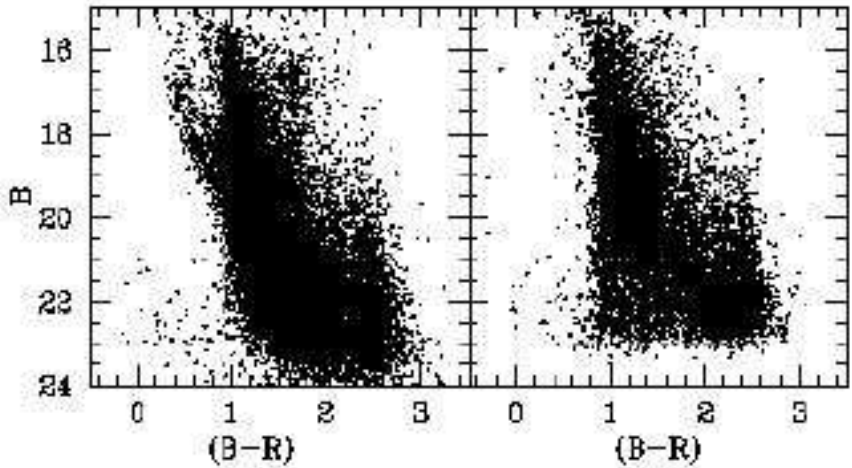}
  \caption{({\it left panel}) Color-magnitude diagram of the center of the Canis Major overdensity ((l,b)=(240,-8)) (see Mart\'\i nez-Delgado et al. 2005 for a detailed
description); ({\it right panel}) Color-magnitude diagram of a control field situated at (l,b)=(240,+8), showing the distribution of Galactic background/foreground stars and an
absence of the dwarf galaxy features.}\label{fig:contour}
\end{figure}

\section{The nature of the Canis Major dwarf }\label{sec:canis}

Fig 2a shows the color-magnitude diagram (CMD) of the putative center of the CMa over-density discovered by Martin et al. (2004). The
most remarkable feature is a narrow main-sequence (MS), with a high constrast
 with respect to the thin/thick/halo background contamination, of which distribution is better observed in
the CMD of a control field (Fig.2b). The CMD morphology  in Fig. 1a  is in agreement with the expected for a stellar system composed by a mean metal-rich stellar population and that has undergone at least two distinct epochs
 of star formation (the last one only 1-2 Gyr ago). 

From the MS feature, we derive a line-of sight depth for
this system of 0.9$\pm$ 0.3 kpc,  consistent with the interpretation of a remnant of a partially disrupted dwarf satellite (Mart\'\i nez-Delgado et al. 2005). This limited line-of-sight depth is also very difficult to reconciled with the hypothesis of a warped Galactic outer disk viewed in projection, as proposed by Momany et al. (2004) (see also Sec. 3). The derived surface brightness ($\mu_{V,0}= 23.3 \pm 0.1$\,mag)  and absolute magnitude ($M_{V}=-14.5 \pm 0.1$\,mag) of CMa places it in the category of dwarf galaxy in the known size-luminosity relation followed by dwarf galaxies (see Pasquali et al. 2005). Additional evidence on the dwarf nature
of CMa from its orbital motion can be found in Sec. 4. At a distance of 8 kpc, CMa is the closest dwarf galaxy known.

\section{The orbit of the Canis Major dwarf}

The question about whether the CMa dwarf could be the progenitor of the Monoceros tidal stream cannot be answered without a better constrain on the motion of
 this posible satellite.  Recent absolute proper motion measurements of a sample of {\it bona-fide} CMa star members (Dinescu et al. 2005) provide an  accurate determination of the CMa orbit. The orbit has a pericenter of 10$\pm$0.9 kpc and an apocenter of 14 $\pm$0.2 kpc. The orbit inclination is 15 $\pm$3 deg and the eccentricity is 0.14$\pm$ 0.04. Currently, CMa is at its apocenter, and it should undergo tidal disruption.

The orbital motion of the CMa also provides important clues on the controversy 
about the origin of this stellar system (dwarf galaxy {\it versus} Galactic pertubation). While the CMa orbit is not very dissimilar from orbits of
thick disk stars, the Monoceros stream stars reach a maximum distance from the Galactic plane of $\sim$ 2 kpc, that is larger than the thick disk scale
height ($\leq 1$ kpc). In addition, the derived W velocity component (W=-49$\pm$15 km/s) shows significant (3$\sigma$) motion perpendicular to the the disk and its negative value is inconsistent (7$\sigma$) with the expected motion (e.g., Drimmel, Smart \& Lattanzi 2000) 
of the warp at this Galactic location (see Dinescu et al. 2005).

The derived orbit parameters are fairly 
similar to those predicted in the theoretical model by Pe\~narrubia et al. (2005; see Sec. 3) for the Monoceros stream's progenitor (within 1$\sigma$).  Fig.1
shows a comparison between the best fitting N-body model (small dots) of the
Monoceros stream and the
orbit of CMa derived from its proper motion (solid line). This agreement supports the argument that CMa is the best 
candidate for the parent galaxy of the Monoceros stream.



\begin{discussion}

\discuss{Mateo}{a)Was the Galactic potential in the model static?; b) How long
did the model run?; c)What is the total mass of the contents of the stream?}

\discuss{Mart\'\i nez-Delgado}{a) In our simulations, the Galaxy reacts to the presence of the Monoceros stream, but our
Galaxy model does not implement a cosmological evolution; b) The available observational data can be reproduce within the last
3 Gyr of a possible progenitor orbit; c) The initial mass was $6 \times 10^8 
M_\odot$ and the satellite lost half of the mass at the end of the simulation.}

\end{discussion}

\end{document}